\documentclass[twocolumn,prb,aps,superscriptaddress,showpacs,floatfix]{revtex4-2}
\usepackage{graphicx}
\usepackage{amssymb,amsmath,mathtools,physics,siunitx}
\usepackage{float}
\usepackage{epsfig}
\usepackage{color}
\usepackage[colorlinks=true, letterpaper=true, pdfstartview=FitV, linkcolor=blue, citecolor=blue, urlcolor=blue]{hyperref}
\usepackage{bm}
\usepackage{multirow}
\usepackage[dvipsnames]{xcolor}
\definecolor{mypurple}{RGB}{112, 48, 160}
\definecolor{myblue}{RGB}{0, 112, 192}
\usepackage{booktabs}
\usepackage{soul}

\begin{document}
\title{Orbital enhanced intrinsic nonlinear planar Hall effect for probing topological phase transition in CuTlSe$_{2}$}

\author{Fan Yang}
\affiliation{School of Physics, Beihang University, Beijing 100191, China}

\author{Xu-Tao Zeng}
\affiliation{School of Physics, Beihang University, Beijing 100191, China}

\author{Huiying Liu}
\email{liuhuiying@buaa.edu.cn}
\affiliation{School of Physics, Beihang University, Beijing 100191, China}

\author{Cong Xiao}
\email{congxiao@fudan.edu.cn}
\affiliation{Interdisciplinary Center for Theoretical Physics and Information Sciences (ICTPIS), Fudan University, Shanghai 200433, China}

\author{Xian-Lei Sheng}
\email{xlsheng@buaa.edu.cn}
\affiliation{School of Physics, Beihang University, Beijing 100191, China}
\affiliation{Peng Huanwu Collaborative Center for Research and Education, Beihang University, Beijing 100191, China}

\author{Shengyuan A. Yang}
\affiliation{Research Laboratory for Quantum Materials, IAPME, University of Macau, Taipa, Macau, China}

\begin{abstract} 
The intrinsic nonlinear planar Hall effect proposed in recent studies 
offers a new way to probe intrinsic band geometric properties in a large class of materials. 
However, the search of material platforms with a large response remains a problem. Here, 
we suggest that topological Weyl semimetals can host enhanced intrinsic nonlinear planar Hall effect. 
From a model study, we show that the enhancement is mainly from the orbital contribution, and the response coefficient exhibits
a characteristic resonance-like lineshape around the Weyl-point energy. 
Using first-principles calculations, we confirm these features for the concrete material CuTlSe$_{2}$. Previous studies have reported two different topological states of CuTlSe$_{2}$. We find this difference originates from two slightly different structures with different lattice parameters. We show that the nonlinear planar Hall response is much stronger in the Weyl semimetal state than in the topological insulator state, and the large response is indeed dominated by orbital contribution amplified by Weyl points.
Our work reveals a close connection between nonlinear orbital responses and topological band features, and suggests CuTlSe$_{2}$ as a suitable platform for realizing enhanced nonlinear planar Hall effect.
\end{abstract}

\maketitle

\section{introduction}
Intrinsic responses, which are unaffected by disorder scattering and constituent inherent characteristics of each material system, have been a focus in condensed matter physics research. Intrinsic responses can be unambiguously and quantitatively evaluated from first-principles calculations, which provides important benchmarks for analyzing experimental data.
Moreover, intrinsic response coefficients usually encodes quantum geometry of the band structure. As a prominent example, the intrinsic anomalous Hall effect is due to the Berry curvature of band structure~\cite{jungwirth2002anomalous, onoda2002topological, nagaosa2010anomalous}.
Recent studies showed that intrinsic responses also occur in nonlinear regime, and these intrinsic nonlinear responses open new possibility to explore a variety of band geometric quantities~\cite{gao2014field,wang2021intrinsic,liu2021intrinsic,xiao2022intrinsic,huang2023intrinsic, huang2023nonlineara,wang2024orbital}.
The first example in this series is the intrinsic nonlinear anomalous Hall effect~\cite{gao2014field, wang2021intrinsic, liu2021intrinsic},
which was shown to arise from the Berry-connection polarizability (BCP)~\cite{liu2022berry}
and can be further connected to the quantum metric of Bloch bands~\cite{wang2021intrinsic, gao2023quantum, feng2024quantum}.

Intrinsic responses have also been proposed in planar Hall transport. For planar Hall measurement, a magnetic field $B$ is applied in the transport plane formed by the driving $E$ field and the response Hall current $j_H$. It was found that
intrinsic contribution may exist and even dominate in linear planar Hall effect, but it has stringent constraints on the crystalline symmetry of the system~\cite{wang2024orbital}. On the other hand, the nonlinear planar Hall effect (NPHE),
where $j_H\propto E^2B$, has less symmetry constraints and hence can be found in a wide range of materials. Its signals have been reported in experiments on several systems, such as the films of Bi$_2$Se$_3$ and SrTiO$_3$~\cite{he2019nonlinear}, CoSi~\cite{esin2021nonlinear}, SrIrO$_3$~\cite{kozuka2021observation,lao2022anisotropic}, WTe$_2$~\cite{he2019nonlinear,yokouchi2023giant}, and Te chiral crystals~\cite{niu2023tunablea}. Early studies focused on an extrinsic mechanism, which gives a response $\propto \tau^2$ ($\tau$ is the scattering time)~\cite{he2019nonlinear, yokouchi2023giant}.
The intrinsic mechanism of NPHE was proposed recently by Huang \emph{et al.}~\cite{huang2023intrinsic, huang2023nonlineara}, for which the response coefficient involves geometric quantities like BCP and its magnetic susceptibility. Notably, for both intrinsic linear and nonlinear planar Hall effect, the in-plane $B$ field acts via a Zeeman coupling, and it was shown that this involves not only spin but also orbital magnetic moment~\cite{huang2023intrinsic,wang2024orbital,pan2025intrinsic}, which could be important in 3D bulk materials. So far, the experimentally studied materials, like Bi$_2$Se$_3$ and WTe$_2$~\cite{he2019nonlinear}, appear to have weak intrinsic response; and conclusive experimental evidence of intrinsic NPHE remains elusive. Hence, it is still an important task to identify suitable material systems for the research of intrinsic NPHE.

CuTlSe$_{2}$ is a representative of a large class of I-III-VI$_2$ ternary chalcopyrite materials~\cite{hahn1953untersuchungen}. This class of materials typically have low cost, low lattice thermal conductivity, can be easily doped (both $n$- and $p$-types), and can form high-quality interfaces with mainstream semiconductors. They have been long studied for thermoelectric, photovoltaic, and light-emitting device applications~\cite{shay1975ternary}. In recent years, there is renewed interest in these materials because of their nontrivial topological character. Particularly, for CuTlSe$_{2}$, in 2011, Feng \emph{et al.}~\cite{feng2011threedimensional} predicted that it is a nontrivial $\mathbb{Z}_2$ topological insulator; later, in 2016, Ruan \emph{et al.}~\cite{ruan2016ideal} proposed that it is an ideal Weyl semimetal. Both works are using first-principles calculations, and the structures studied in the two works differ slightly in the lattice constants. A very recent experiment on CuTlSe$_{2}$, via magneto-transport and magnetic characterizations, appeared to support the Weyl semimetal prediction~\cite{wang2024weyl}.

In this work, we investigate the intrinsic NPHE in CuTlSe$_{2}$, based on model and first-principles calculations. We find that this material actually undergoes a topological phase transition {determined by the structure's tetragonal distortion}, which explains the different results in previous works. We show that NPHE tends to be dramatically enhanced by Weyl points in band structure, and it features a characteristic resonance-like lineshape across the Weyl-point energy.
Through a model study, we reveal the enhancement is largely due to the orbital contribution, which is sensitive to band crossing points.
These features can be clearly observed in the Weyl semimetal state of CuTlSe$_{2}$, which exhibits a much stronger intrinsic NPHE than in the topological insulator state. Our work reveals CuTlSe$_{2}$ as a promising platform for realizing strong intrinsic NPHE and for studying its interplay with band topology.
The revealed high sensitivity of NPHE response to band topology also offers a new tool for
probing topological phase transitions.

\section{formulation of intrinsic NPHE}

Let's first briefly review the theory of intrinsic NPHE. The key physical quantity of interest is the NPHE response tensor $\chi$, which is defined from the relation~\cite{huang2023intrinsic, huang2023nonlineara}
\begin{equation}\label{rel}
    j_a = \chi_{abcd} E_b E_c B_d,
\end{equation}
where the subscripts label the Cartesian components. Consider the configuration where the transport plane is the $x$-$y$ plane, then all subscripts in Eq.~(\ref{rel}) belong to $\{x,y\}$.

For intrinsic contribution, the corresponding $\chi$ tensor is solely determined by the band structure of the material system.
The formula has been derived in Ref.~\cite{huang2023intrinsic, huang2023nonlineara} based on the extended semiclassical theory~\cite{gao2014field,gao2015geometricala} (we set $e=\hbar=1$ and assume the system is 3D):
\begin{equation}
\begin{aligned}
    \chi_{abcd}
    =&\sum_n \int \frac{d^3 k}{(2 \pi)^3} f_0^{\prime}(\varepsilon_n)\Big[\left(v_a^n \Lambda_{b c d}^n-v_b^n \Lambda_{a c d}^n\right) \\
    &\qquad +\left(\partial_a G_{b c}^n-\partial_b G_{a c}^n\right) \mathcal{M}_{d}^{n}\Big].
\end{aligned}\label{Eq:chi_abcd}
\end{equation}
Here, we have suppressed explicit $k$ dependence of quantities in the integrand, $f_0$ is the equilibrium Fermi distribution, $n$ is the band index,
$\partial_a\equiv \partial/\partial k_a$, $v^n_a$ is the usual band velocity, and $\mathcal{M}^{n}_d$ is the intraband magnetic moment (its expression is given in Eqs.~(\ref{MM}-\ref{Mo}) below).
$G$ is the BCP tensor~\cite{liu2022berry}, expressed as
\begin{equation}
    G^n_{a b}=2 \operatorname{Re} \sum_{m \neq n} \frac{v_a^{n m} v_b^{m n}}{\left(\varepsilon_n-\varepsilon_m\right)^3},
\end{equation}
where $v^{nm}_a$ is the interband velocity matrix element.
And $\Lambda$ is the magnetic susceptibility of BCP~\cite{huang2023intrinsic, huang2023nonlineara}, i.e., it gives the correction to BCP due to applied in-plane $B$ field. Its expression is given by
\begin{equation}
\begin{aligned}
    \Lambda_{abc}^n(k)=2 \operatorname{Re} \sum_{m \neq n}&\left[\frac{3 v_a^{n m} v_b^{m n}\left(\mathcal{M}_{c}^{n }-\mathcal{M}_{c}^{ m}\right)}{\left(\varepsilon_n-\varepsilon_m\right)^4}\right. \\
    -\sum_{l \neq n} &\frac{\left(v_a^{l m} v_b^{m n}+v_b^{l m} v_a^{m n}\right) \mathcal{M}_{c}^{n l}}{\left(\varepsilon_n-\varepsilon_l\right)\left(\varepsilon_n-\varepsilon_m\right)^3} \\
    -\sum_{l \neq m} &\left.\frac{\left(v_a^{l n} v_b^{n m}+v_b^{l n} v_a^{n m}\right) \mathcal{M}_{c}^{m l}}{\left(\varepsilon_m-\varepsilon_l\right)\left(\varepsilon_n-\varepsilon_m\right)^3}\right],
\end{aligned}\label{Eq:BCP}
\end{equation}
where $\bm{\mathcal{M}}^{nl}$ is the interband matrix element of magnetic moment.

$\mathcal{M}$ contains two parts:
\begin{equation}\label{MM}
  \mathcal{M}=\mathcal{M}_S+\mathcal{M}_O.
\end{equation}
$\mathcal{M}_S$ stands for the spin part, with
\begin{equation}\label{Ms}
   \bm{\mathcal{M}}_{S}^{mn}=-g \mu_B \bm{s}^{m n},
\end{equation}
where $g$ is the $g$ factor, $\mu_B$ is the Bohr magneton, and $\bm{s}^{mn}$ is the spin matrix element.
Meanwhile, $\mathcal{M}_O$ is the orbital part, with~\cite{pozoocana2023multipole}
\begin{equation}\label{Mo}
    \bm{\mathcal{M}}_{O}^{mn}=\frac{1}{4 i} \sum_{p \neq m, n}\left(\frac{1}{\varepsilon_p-\varepsilon_m}+\frac{1}{\varepsilon_p-\varepsilon_n}\right) \bm{v}^{m p} \times \bm{v}^{p n}.
\end{equation}
The intraband magnetic moment just corresponds to the diagonal matrix element, i.e., $\bm{\mathcal{M}}^n\equiv \bm{\mathcal{M}}^{nn}$.

We have a few remarks regarding these formulas. First, the intrinsic response described by Eq.~(\ref{Eq:chi_abcd}) is a purely Hall response. This can be directly observed by noting that $\chi$ is antisymmetric in its first two indices. It indicates that the resulting current is dissipationless, with $\bm j\cdot\bm E=0$. (One may choose to symmetrize the middle two indices $b$ and $c$, which we did not do here for simpler expressions.)

Second, the orbital magnetic moment $\mathcal{M}_O$ originates from the orbital motion of Bloch electrons. In the planar Hall setup, only the in-plane component of $\mathcal{M}_O$ couples with the $B$ field. One can expect that its effect is less important in 2D systems, as the electron's orbital motion is confined in the out-of-plane direction. Indeed, Ref.~\cite{huang2023intrinsic, huang2023nonlineara} showed that in 2D material MoSSe, the orbital contribution is much smaller than the spin contribution. Nevertheless, the orbital contribution could generally be significant for 3D systems~\cite{wang2024orbital,pan2025intrinsic}.

Third, comparing Eqs.~(\ref{Ms}) and (\ref{Mo}), one can observe that while the spin moment is always bounded, the orbital moment may diverge when approaching band degeneracies. This kind of singular behavior is also seen in band geometric quantities like Berry curvature and BCP. Physically, it represents the strong interband coherence around a band degeneracy point. To understand this important feature, in the following, we consider the response from a Weyl point, which is the most elementary and generic band degeneracy in band structure.

\begin{figure}[t]
  \includegraphics[width=8cm]{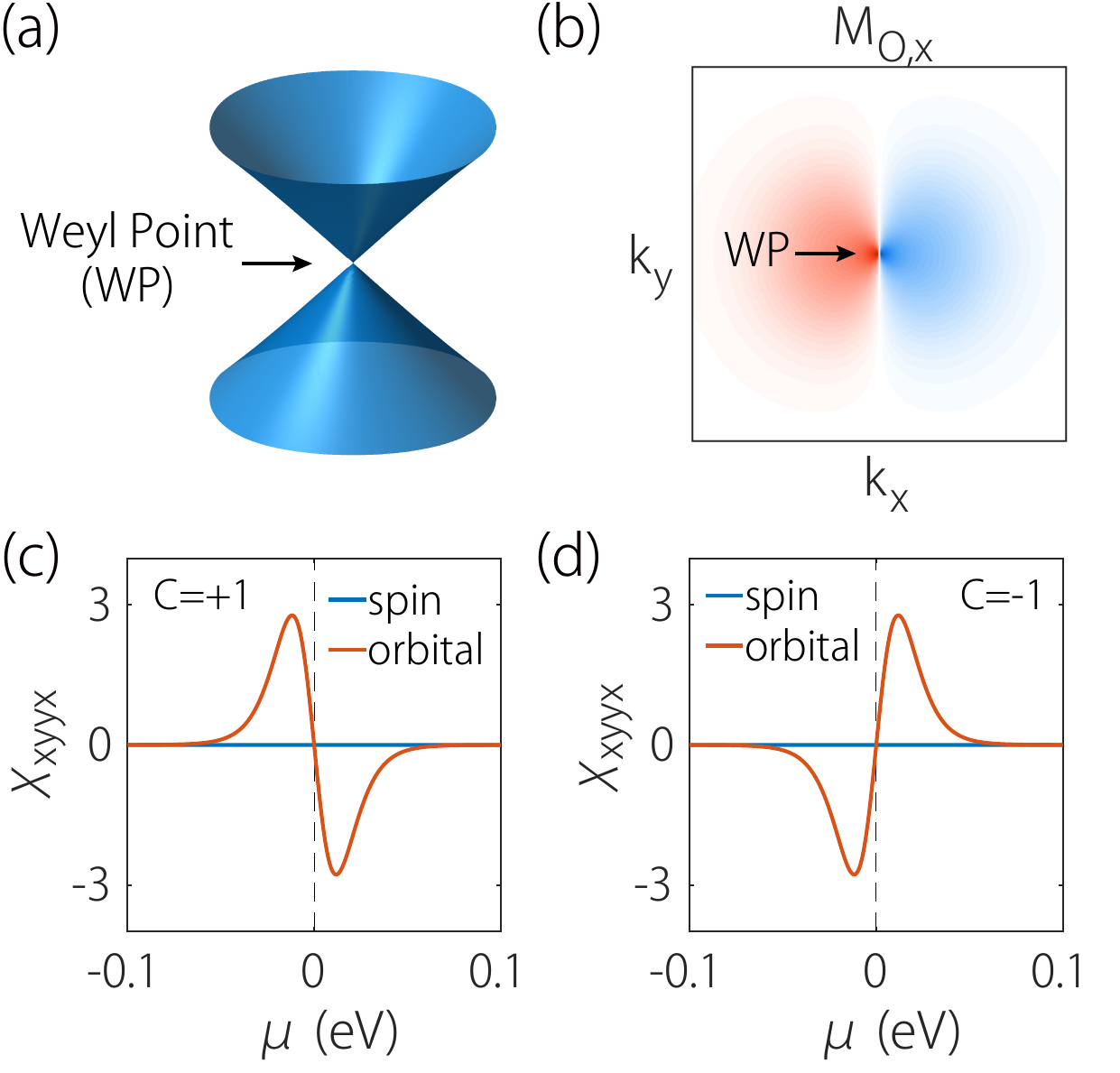}
  \caption{\label{fig:1}
  (a) Schematic of linear band dispersion around a Weyl point (Eq.~(\ref{Hw})). Here, we take the $k_z=0$ plane. 
  (b) The distribution of the orbital magnetic moment $\mathcal{M}_{O,x}^{v}$ for valence band in the $k_z=0$ plane. The Weyl point is indicated with an arrow.
  (c-d) Calculated intrinsic NPHE tensor element $\chi_{xyyx}$ for the Weyl model Eq.~(\ref{Hw}). (c) is for chirality $C=\text{sgn}(v_F)=+1$, and (d) is for chirality $C=-1$. The red (blue) curve is the orbital (spin) contribution. $\chi$ is in unit of $10^{-5}$AT$^{-1}$V$^{-2}$. In the calculation, we take $v_F=1$~eV$\cdot$\AA\ and a temperature of 100~K.
  }
\end{figure}

\section{NPHE from a Weyl point}

Weyl points are accidental band crossing points formed between two bands~\cite{wan2011topological}.
They act like monopoles for Berry curvature fields in momentum space. The semimetal state with Weyl points near the Fermi level is known as
the Weyl semimetal, for which a range of fascinating physical properties have been proposed in the past decade~\cite{armitage2018weyl}.
Here, we consider the intrinsic NPHE contributed by a Weyl point. The simplest model reads

\begin{equation}\label{Hw}
    \mathcal{H}=v_F\bm k\cdot\bm \sigma,
\end{equation}
where the energy and the wave vector $\bm k$ are measured from the Weyl point, $\bm\sigma$ is the vector of Pauli matrices representing spins, and $v_F$ is a real model parameters whose magnitude gives the Fermi velocity.

The spectrum of this model consists of two linearly crossing bands, labeled as $c$ and $v$, with energy dispersion
\begin{equation}
  \varepsilon_{c/v}(\bm k)=\pm |v_F|k
\end{equation}
where $k$ is the magnitude of the wave vector, as shown in Fig.~\ref{fig:1}(a).
The Weyl point, which is the band crossing point at $\bm k=0$, has a chirality/handedness, which is given by $\text{sgn}(v_F)$.

Let's first examine the magnetic moments $\mathcal{M}_S$ and $\mathcal{M}_O$ for electrons near the Weyl point.
According to Eqs.~(\ref{Ms}) and (\ref{Mo}), we find that
\begin{equation}\label{WMs}
  \mathcal{M}_{S,a}^{v}=-\mathcal{M}_{S,a}^{c}=\text{sgn}(v_F)\frac{g\mu_B k_a}{2k},
\end{equation}
\begin{equation}\label{WMo}
  \mathcal{M}_{O,a}^{v}=\mathcal{M}_{O,a}^{c}=-\frac{v_Fk_a}{2k^2},
\end{equation}
where, again, the index $a$ labels Cartesian component of a vector. Notably, $\mathcal{M}_{S}$ has a bounded value, but
$\mathcal{M}_{O}$ tends to diverge as $k^{-1}$ when approaching the Weyl point. Their ratio,
\begin{equation}
  \left|\frac{\mathcal{M}_{S,a}^{p}}{\mathcal{M}_{O,a}^{p}}\right|=\frac{g\mu_B k}{|v_F|},
\end{equation}
with $p=c,v$, shows that the orbital moment will dominate over the spin moment near the Weyl point. In Fig.~\ref{fig:1}(b), we plot the distribution of $\mathcal{M}_{O,x}^v$ in the momentum space. One observes that it exhibits a dipole-like pattern: the dipole is in the $x$ direction and its strength is concentrated around the Weyl point.

\begin{figure}[t]
  \includegraphics[width=8cm]{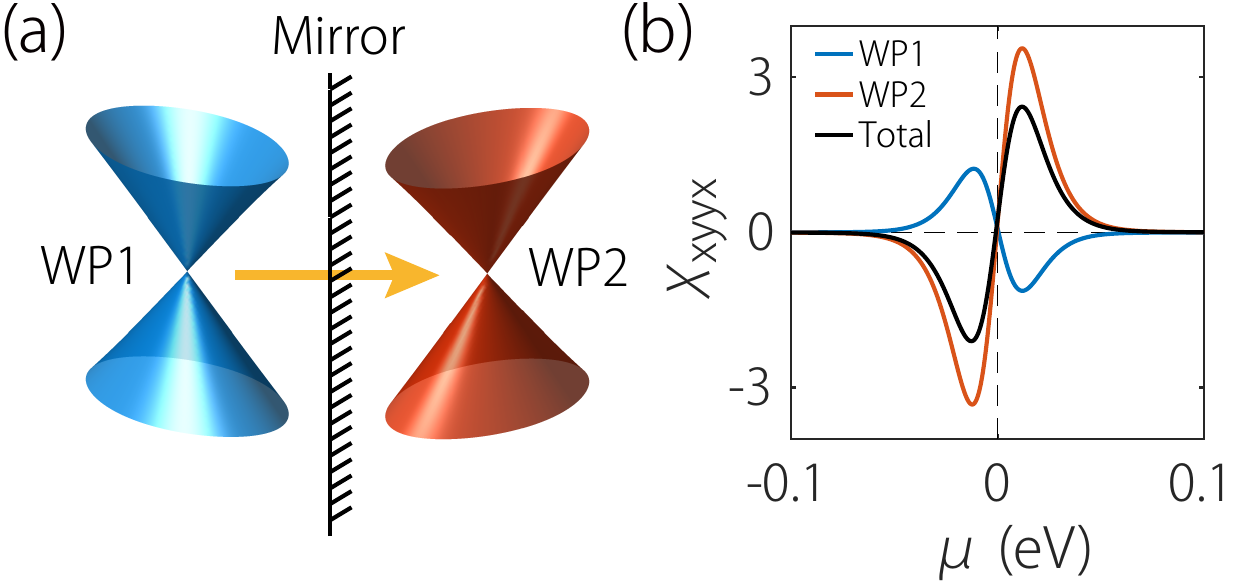}
  \caption{\label{fig:2}
  (a) Schematic figure showing two anisotropic Weyl points, WP1 and WP2, connected by a mirror $M_{xy}$. Due to the symmetry operation, the two points have opposite chirality.
  (b) Intrinsic NPHE tensor element $\chi_{xyyx}$ due to WP1, WP2, and their sum. $\chi$ is in unit of $10^{-5}$AT$^{-1}$V$^{-2}$. In the calculation, we take $v_x=1.4$~eV$\cdot$\AA, $v_y=0.6$~eV$\cdot$\AA, $v_z=1$~eV$\cdot$\AA\ in model (\ref{An}), and a temperature of 100~K.
  }
\end{figure}

Next, we compute the intrinsic NPHE response tensor $\chi$ for this model, using formula Eq.~(\ref{Eq:chi_abcd}). The calculation is straightforward. We find that for this simple model, the spin contribution vanishes identically, and the response is entirely from the orbital contribution, with
\begin{equation}
  \chi_{xyyx}=-\text{sgn}(v_F \mu)\frac{v_F^2}{24\pi^2}\int_0^\infty d\varepsilon\ \frac{f_0'(\varepsilon-|\mu|)}{\varepsilon^3}.
\end{equation}
One finds that the response coefficient has opposite signs for electron-doped ($\mu>0$) and hole-doped ($\mu<0$) cases. And the sign also flips if the chirality of Weyl point is reversed. For $T\rightarrow 0$, the response has a singular behavior at Weyl point energy, which diverges as $\mu^{-3}$. At finite temperature (or with finite level broadening), the singularity is removed and  $\chi_{xyyx}$
exhibits a resonance-like
lineshape when plotted as a function of chemical potential $\mu$, i.e., it shows two peaks with opposite signs across the Weyl point, similar to the behavior of dielectric function around a resonance, as shown in Figs.~\ref{fig:1}(c-d). One observes that overall, the response is concentrated in region around the Weyl point and quickly decays when moving away from the Weyl point energy. These are characteristic features of Weyl-point enhanced NPHE.

It is worth noting that the signs of magnetic moments in Eqs.~(\ref{WMs},\ref{WMo}) and also the response tensor depend on the chirality of Weyl point. This means a Weyl point with $v_F\rightarrow -v_F$  will give the opposite contribution to response $\chi$. According to Nielsen-Ninomiya no-go theorem~\cite{nielsen1981nogo}, the Weyl points appear in pairs of opposite chirality~\footnote{There may appear a case where a single Weyl point coexists with nodal surfaces or other features, as shown in Ref~\cite{yu2019circumventing}. We do not consider such special cases here.}. Does this imply that the overall NPHE response from all Weyl points will be zero? Generally, the answer is no.
For example, if two Weyl points of opposite chirality are sitting at different energies, then their contributions will not cancel out.
And even if the Weyl points are at the same energy, anisotropy may also make a difference.
In the discussion above, we took the simplest Weyl model, Eq.~(\ref{Hw}), which is isotropic. However, Weyl points usually are not located at high-symmetry points of Brillouin zone, and hence the dispersion around a Weyl point is generally anisotropic. Due to this anisotropy, contributions to a particular tensor element from two Weyl points may differ in magnitude, leading to a nonzero net result. For example, consider a Weyl point with
anisotropic dispersion described by
\begin{equation}\label{An}
  H_{W}=v_x k_x\sigma_x+v_y k_y\sigma_y+v_zk_z \sigma_z,
\end{equation}
and another Weyl point $H_{W'}$ with opposite chirality connected to $H_{W}$ by a mirror $M_{xy}$, as schematically shown in Fig.~\ref{fig:2}(a). In Fig.~\ref{fig:2}(b), we plot their individual and total contributions to $\chi_{xyyx}$. One observes that although the two have results with opposite signs, the magnitudes are different, so we are still left with a non-vanishing result with the characteristic resonance-like peak feature around the energy of Weyl points.

\begin{figure}[t!]
    \includegraphics[width=8cm]{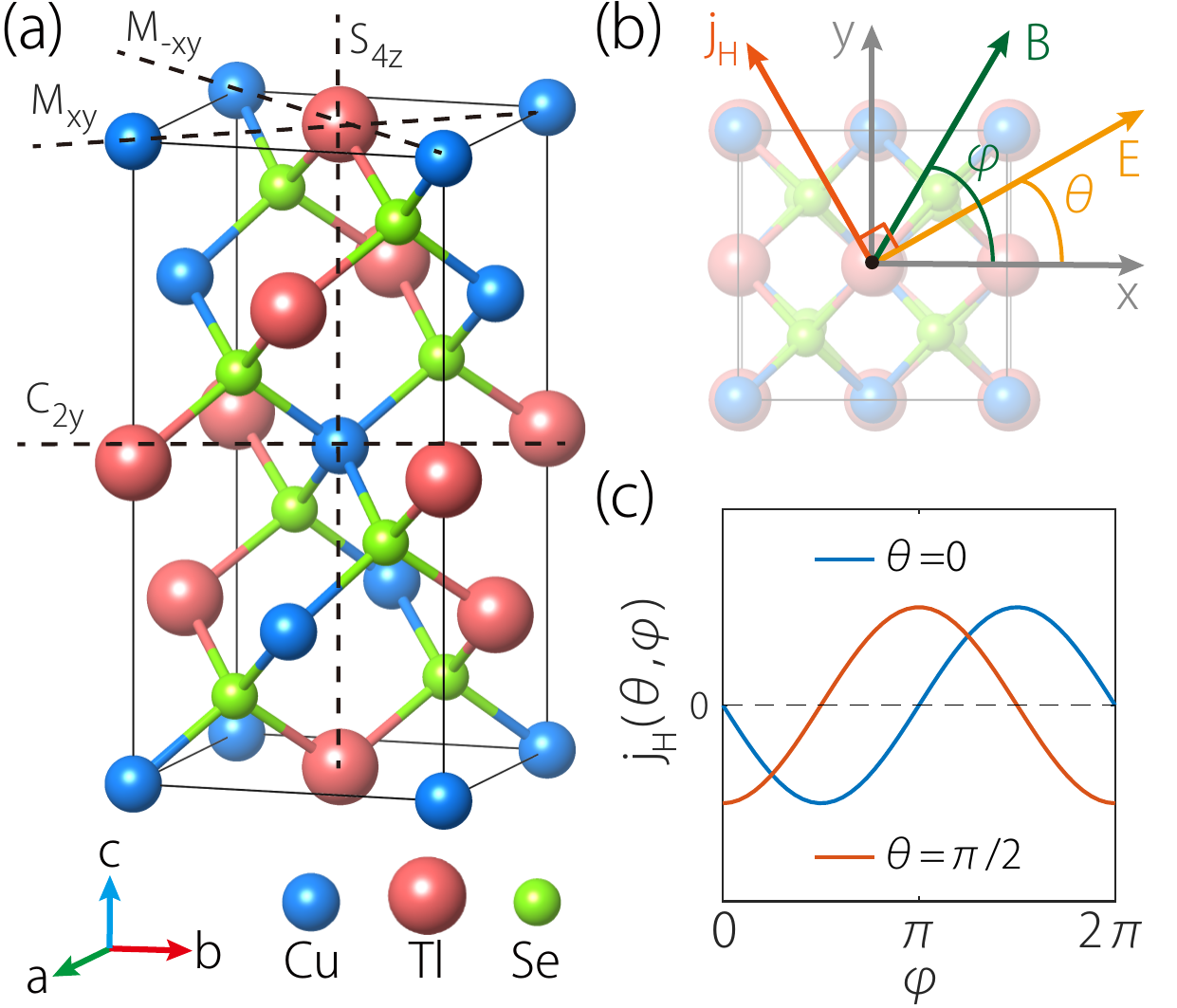}
    \caption{\label{fig:3}
    (a) Crystal structure of CuTlSe$_{2}$. The solid black lines mark the conventional cell.
    Some of the point-group symmetries are indicated in the figure.
    (b) Schematic of the measurement setup for NPHE. We take the crystal $a$ axis as $x$ direction and $b$ axis as $y$ direction. The in-plane $E$ and $B$ fields are specified by angles $\theta$ and $\varphi$, respectively. And the Hall current is in the direction normal to $E$ field.
    (c) Schematic of the angular dependence of intrinsic NPHE. The two curves correspond to cases with $E$ field aligned with the crystal $a$ axis ($\theta=0$) and $b$ axis ($\theta=\pi/2$), respectively.}
\end{figure}

\section{Application to $\text{CuTlSe}_2$}

As mentioned, CuTlSe$_2$ belongs to the large family of I-III-VI$_2$ chalcopyrite compounds, which are isoelectronic analogs of II-VI binary semiconductors. Its crystal structure is illustrated in Fig.~\ref{fig:3}(a). It can be regarded as a superlattice of zinc-blende structure with small structural distortions, i.e., its unit cell doubles that of the cubic zinc-blende unit cell in the $c$ direction. Cu and Tl are ordered on two different cation sites, and each anion Se site is surrounded by a tetrahedron formed by two Cu and two Tl atoms. The structure of chalcopyrite compounds generally exhibits a tetragonal distortion, which is characterized by the ratio of lattice constants $\eta=c/2a$.

Let's first examine the symmetry constraints on the intrinsic NPHE response.
The space group of CuTlSe$_{2}$ is $I\bar{4}2d$ (No.~122). Its point group is $D_{2d}$, which contains  two rotations $C_{2x}$ and $C_{2y}$,
two vertical mirrors $M_{xy}$ and $M_{-xy}$,
and a rotoreflection $S_{4z}$. These operations are schematically marked in Fig.~\ref{fig:3}(a). To study NPHE, we consider the $x$-$y$ plane as the transport plane. Among the four independent elements $\chi_{xyyy}$, $\chi_{yxxx}$, $\chi_{xyyx}$, and $\chi_{yxxy}$, we find that $C_{2x}$ and $C_{2y}$ require that $\chi_{xyyy}=\chi_{yxxx}=0$; the three symmetries
$M_{xy}$, $M_{-xy}$, and $S_{4z}$ connect $\chi_{xyyx}$ and $\chi_{yxxy}$ with
\begin{equation}
  \chi_{xyyx}=-\chi_{yxxy}.
\end{equation}
Therefore, there is only one nonzero independent tensor element for intrinsic NPHE, and we take it to be $\chi_{xyyx}$. We also note that this symmetry forbids the linear planar Hall effect $\propto EB$, making NPHE the leading order planar Hall response.

%\hl{The symmetry constraints will further forbid all the linear planar Hall effects in CuTlSe$_{2}$. As for the time-reversal-odd ($T$-odd) part like $j_H \propto \tau EB$, it is not allowed in our nonmagnetic system. And for the $T$-even part like $j_H \propto EB$, the in-plane response current will be forbidden by $S_{4z}$. So the intrinsic NPHE can be the dominant effect among the planar Hall effects in CuTlSe$_{2}$.
%}

In experimental detection, the  $E$ and $B$ fields may not be aligned with the crystal axis. Specify their in-plane directions by angles $\theta$ and $\varphi$ measured from the crystal $a$ axis (taken as $x$ direction), as shown in Fig.~\ref{fig:3}(b). The response Hall current $\bm j_H$ is in the direction $(\hat z\times \bm E)$, with a value
\begin{equation}
   j_H=\bm j_H\cdot (\hat z\times \hat{ \bm E}),
\end{equation}
where hat denotes unit vectors.
The corresponding effective NPHE conductivity is
\begin{equation}
  \chi_H=\frac{j_H}{E^2 B}.
\end{equation}
Expressing $\chi_\text{NPHE}$ in terms of $\chi_{xyyx}$, we obtain a relation with a simple angular dependence:
\begin{equation}\label{angle}
  \chi_H(\theta,\varphi)=-\chi_{xyyx}\sin(\theta+\varphi),
\end{equation}
namely, $\chi_H$ depends on the directions of $E$ and $B$ fields only through the combination {of $(\theta+\varphi$) (see Fig.~\ref{fig:3}(c)).}
This behavior can be directly verified in experiment by using multiple-lead measurement on a disk-shaped sample.
And Eq.~(\ref{angle}) will also be useful for extracting $\chi_{xyyx}$ from experimental data when rotating the in-plane $B$ field direction.

As mentioned, previous studies proposed different topological phases of CuTlSe$_{2}$. We note that this difference originates from the slightly different lattice constants of CuTlSe$_{2}$ crystal. As summarized in Table~\ref{tab:1}, an early experiment in 1953~\cite{hahn1953untersuchungen} (referred to as Exp1953 below) obtained CuTlSe$_{2}$ samples with $a=5.844$ \AA\, and $c=11.623$ \AA; whereas a more recent experiment~\cite{wang2024weyl} (referred to as Exp2024 below) reported CuTlSe$_{2}$ samples with $a=5.822$ \AA\, and $c=11.741$ \AA. The main difference between the two reported structures is in the degree of tetragonal distortion. The samples in Exp1953 has $\eta\approx 0.99$, while those in  Exp2024 has $\eta\approx 1.01$.
The theoretical study in Ref.~\cite{feng2011threedimensional} adopted the CuTlSe$_{2}$ structure parameters in Exp1953, which gives a $\mathbb{Z}_2$ topological insulator state. In comparison, the calculation in Ref.~\cite{ruan2016ideal} is using a structure that is more close to that of Exp2024, leading to a Weyl semimetal state.

\begin{table}[t]
  \caption{\label{tab:1}
  Lattice parameters of CuTlSe$_2$ reported from two experimental studies, Ref.~\cite{hahn1953untersuchungen} and Ref.~\cite{wang2024weyl}. The corresponding band gaps are from our calculations.}
  \begin{ruledtabular}
  \begin{tabular}{lcc}
  \textrm{}&
  \textrm{Ref.~\cite{hahn1953untersuchungen}}&
  \textrm{Ref.~\cite{wang2024weyl}}\\
  \colrule
  Lattice $a$ (\AA) & 5.844  &  5.822 \\
  Lattice $c$ (\AA) & 11.623 & 11.741 \\
  Band gap (meV)    & 10.3   & 0.0 \\
  Topology         & Topological insulator & Weyl semimetal \\
  \end{tabular}
  \end{ruledtabular}
\end{table}

\begin{figure}[tbp]
  \includegraphics[width=8cm]{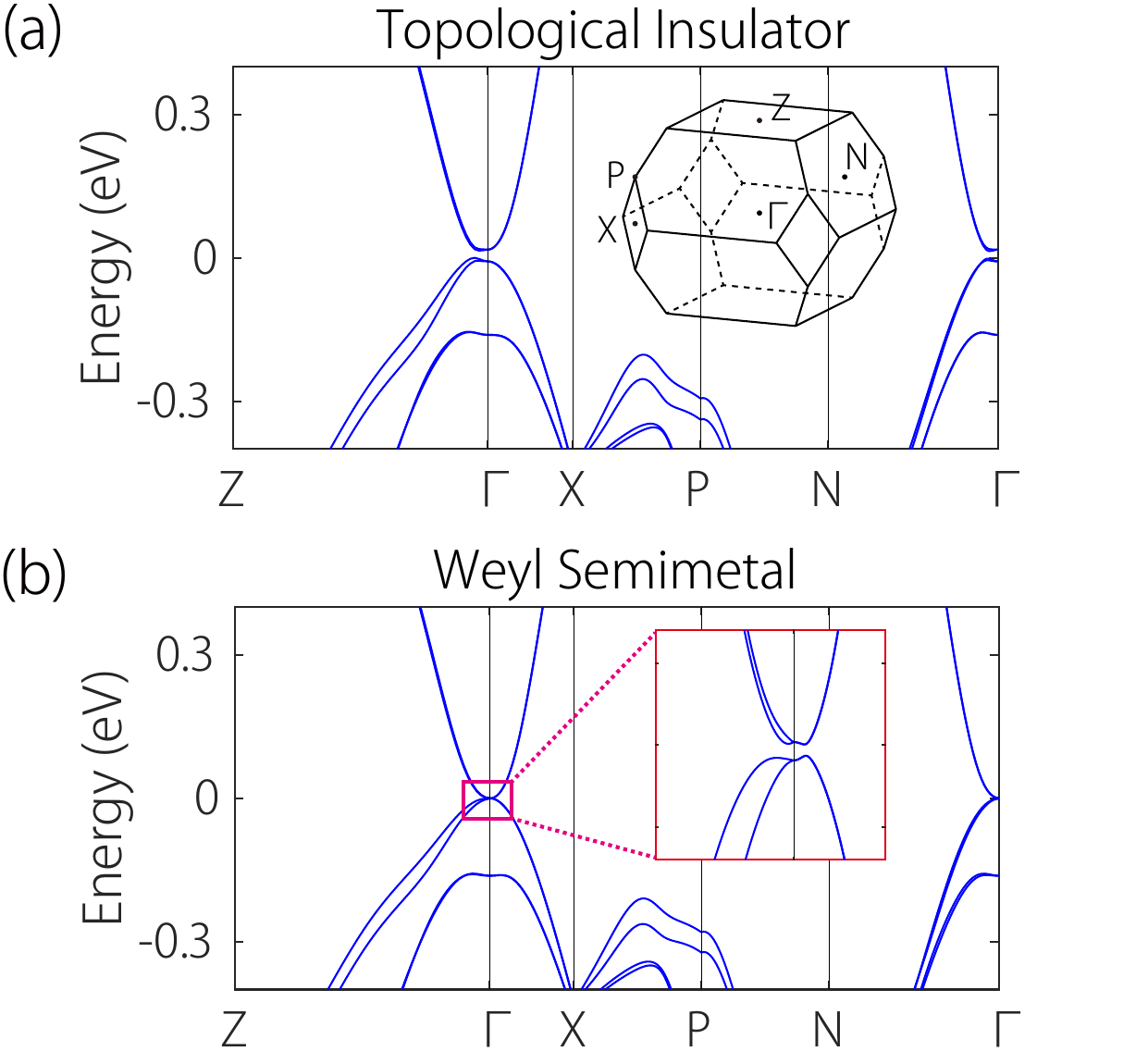}
  \caption{\label{fig:4} The two topological states of CuTlSe$_{2}$.
  (a) Band structure computed using the crystal structure of Ref.~\cite{hahn1953untersuchungen}, which is a topological insulator.
  (b) Band structure computed using the crystal structure of Ref.~\cite{wang2024weyl}, which is a Weyl semimetal. The Weyl points are off the high symmetry paths (see next figure).
  }
\end{figure}

\begin{figure}[tbp]
  \includegraphics[width=8cm]{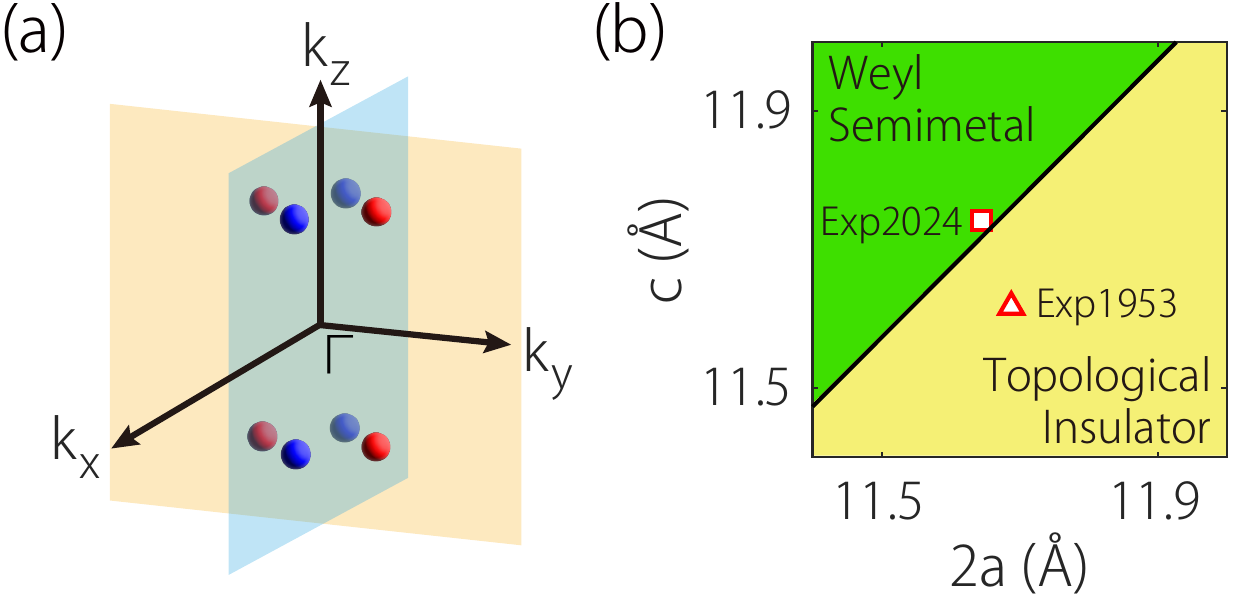}
  \caption{\label{fig:5}
  (a) Schematic illustration of positions of the eight Weyl points of CuTlSe$_{2}$ in Weyl semimetal state. They are located  in the $k_x$-$k_z$ and $k_y$-$k_z$ planes of the Brillouin zone.
  The Weyl points with chirality $C=\pm1$ are marked with red and blue colors, respectively.
  (b) Phase diagram of CuTlSe$_{2}$
  with respect to the lattice constants $a$ and $c$.
  The two experimentally reported structures are marked.}
\end{figure}

We perform first-principles calculations (see Appendix for details of calculation) on the two experimentally reported CuTlSe$_{2}$ structures and obtain results that are consistent with the previous studies. In Figs.~\ref{fig:4}(a-b), we plot the computed low-energy band structures for the two structures. One observes that in the Exp1953 structure, CuTlSe$_{2}$ is a narrow-gap semiconductor, with a gap value about 10 meV.
Evaluating its $\mathbb{Z}_2$ topological index gives a nontrivial value, indicating the system is a topological insulator,
consistent with the finding of Ref.~\cite{feng2011threedimensional}. Meanwhile, for the result with Exp2024 structure in Fig.~\ref{fig:4}(b), although the spectrum {along high-symmetry paths} appears to have a small band gap, a careful scan of low-energy bands reveals that there exist crossing points between conduction and valence bands in two constant $k_z$ planes with $k_z=\pm k_z^*=0.0091$ \AA$^{-1}$. These are the Weyl points, and there are totally eight of them, as illustrated in Fig.~\ref{fig:5}(a). Interestingly, the four Weyl points in the $k_x$-$k_z$ plane have negative chirality, whereas the other four in the $k_y$-$k_z$ plane have positive chirality. All the eight Weyl points are connected by symmetry, and there is no other extraneous bands around. Hence, for a system without doping, the Weyl points must be located at the Fermi level, enforced by band filling. This is why it was referred to as an ideal Weyl semimetal~\cite{ruan2016ideal}.

Having seen that the difference in band topology arises from difference in lattice constants, there should exists a topological phase transition driven by lattice strain. In Fig.~\ref{fig:5}(b), we show a calculated phase diagram of CuTlSe$_{2}$. One observes that a phase boundary exists at {$\delta\equiv (c-0.073)/2a\approx 1$, where $\delta<1$ indicates a topological insulator phase and $\delta>1$ indicates a Weyl semimetal phase.} A compressive (tensile) strain applied along the $c$ axis tends to drive the system towards topological insulator (Weyl semimetal) phase.

\begin{figure}[tbp]
  \includegraphics[width=8cm]{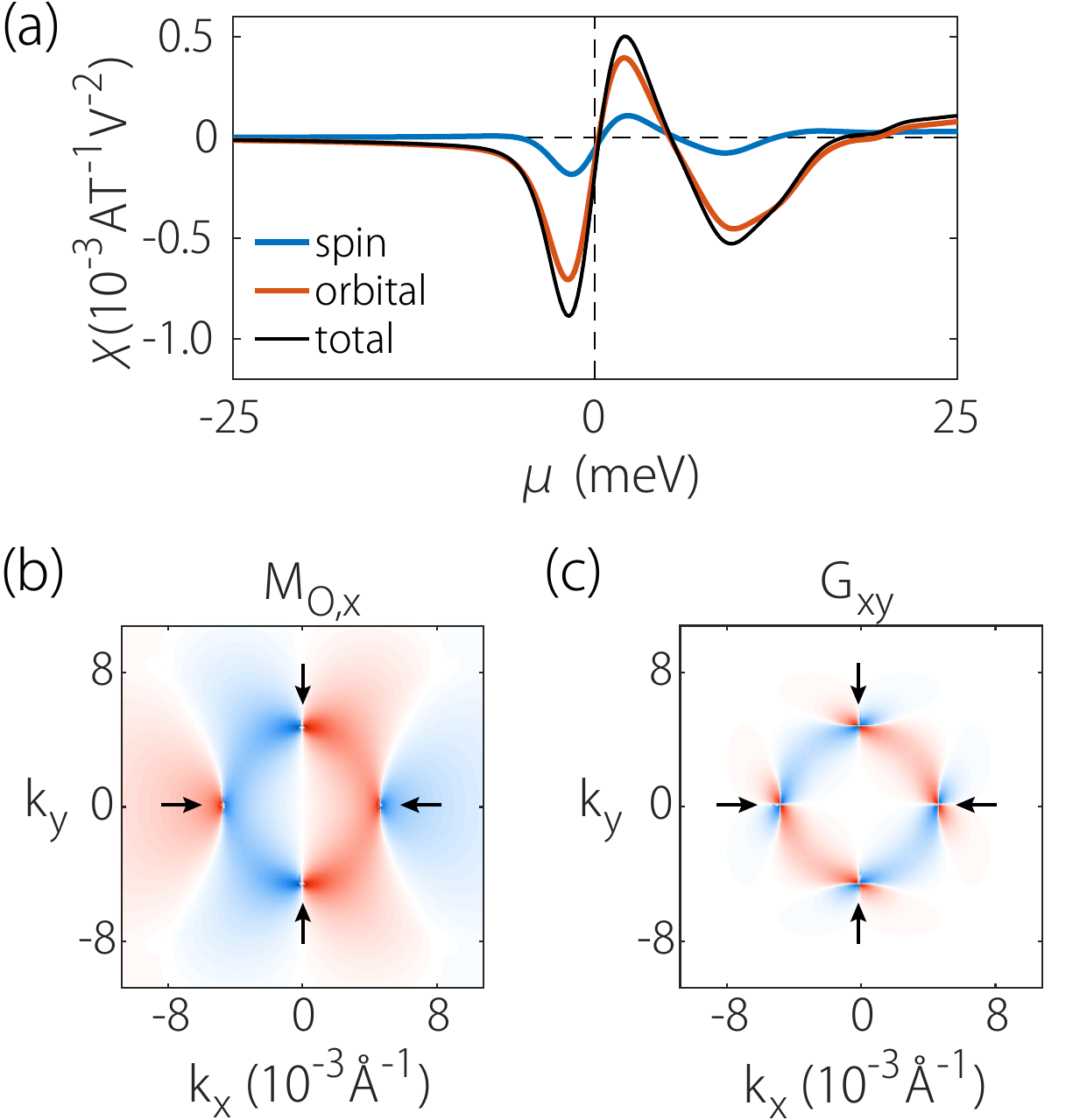}
  \caption{\label{fig:6}
  (a) Calculated intrinsic NPHE tensor
  element $\chi_{xyyx}$ of CuTlSe$_{2}$ in the Weyl semimetal state. The spin and the orbital contributions are also separately plotted.
  (b) Distribution of the orbital magnetic moment $\mathcal{M}_{O,x}$ in $k_z=k_z^*$ plane. 
  (c) Distribution of the BCP tensor $G_{xy}$ in $k_z=k_z^*$ plane. 
 In (b) and (c), the four black arrows indicate the positions of the four Weyl points.
  }
\end{figure}

Based on the first-principles band structure results, we evaluate the intrinsic NPHE response for the two states. First, consider the Weyl semimetal state in Fig.~\ref{fig:4}(b). The calculated NPHE response tensor element $\chi_{xyyx}$ is plotted in Fig.~\ref{fig:6}(a), as a function of chemical potential.
Importantly, it is obvious that the curve shows the characteristic resonance-like lineshape around the Weyl-point energy ($\mu=0$), similar to that in Fig.~\ref{fig:1}(b). {The peak value can reach $\sim 1\times 10^{-3}$ AT$^{-1}$V$^{-2}$. This value is pronounced compared to} the intrinsic NPHE found in 2D MoSSe~\cite{huang2023intrinsic, huang2023nonlineara}.
There is another peak at $\mu \approx 9$ meV, which is associated to {a band crossing in the conduction band}.
In Fig.~\ref{fig:6}(a), we also separately plot the orbital and spin contributions to $\chi_{xyyx}$. It is evident that around the orbital contribution is much larger than the spin contribution, due to the significant enhancement by Weyl points.

To confirm the important role of Weyl points in enhancing NPHE, in Figs.~\ref{fig:6}(b-c), we plot the $k$-space distribution of orbital moment $\mathcal{M}_{O,x}$ and the BCP tensor element $G_{xy}$ in the $k_z=k_z^*$ plane. One can see that all these quantities are enhanced at regions around the Weyl points. Particularly, near each Weyl point, they exhibit the characteristic patterns as we have seen in the model study. For example, the orbital moment distribution near a Weyl point in Fig.~\ref{fig:6}(b) shows the dipole-like pattern, similar to that in Fig.~\ref{fig:1}(b).

\begin{figure}[tbp]
  \includegraphics[width=8cm]{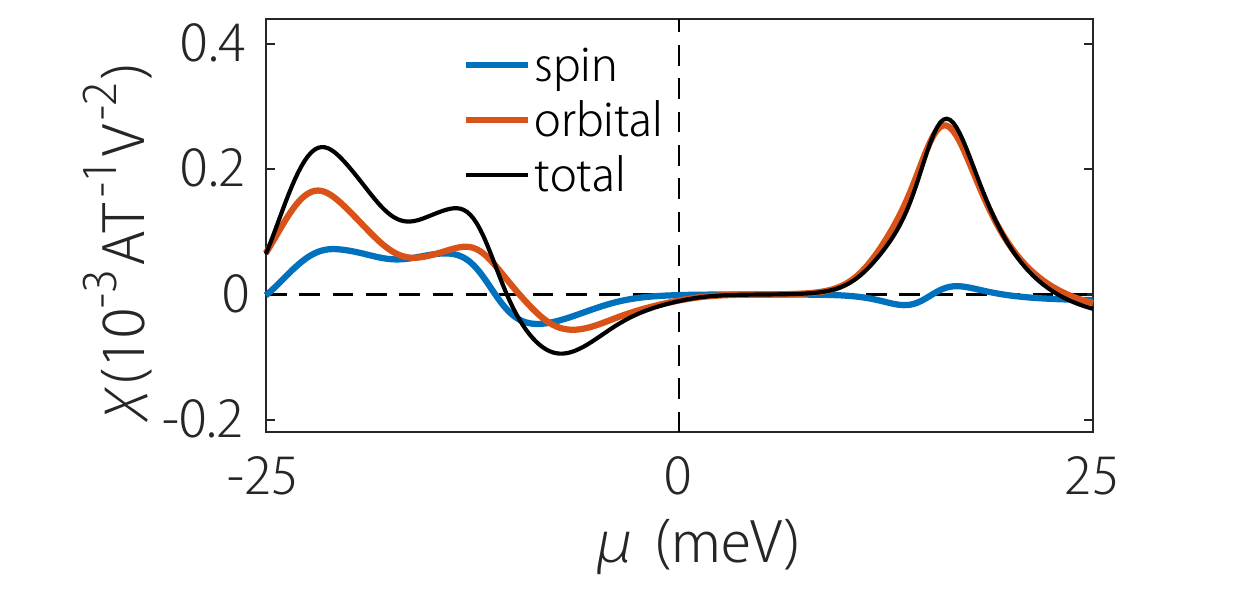}
  \caption{\label{fig:7}
  Calculated intrinsic NPHE tensor
  element $\chi_{xyyx}$ of CuTlSe$_{2}$ in the topological insulator state. The spin and the orbital contributions are also separately plotted.
  }
\end{figure}

Next, we consider the NPHE response for the topological insulator state in Fig.~\ref{fig:4}(a). The calculation results are shown in Fig.~\ref{fig:7}.  One can see that the response still shows peaks at band edges, since the geometric quantities and density of states are in general larger in such regions. However, the values of these peaks are suppressed compared to the Weyl semimetal state. Experimentally, it was found that CuTlSe$_{2}$ samples tend to be hole-doped~\cite{wang2024weyl}. {Comparing the peaks on the hole side in} Fig.~\ref{fig:6}(a) and Fig.~\ref{fig:7}, { one finds that the peak value ($\sim 0.1\times 10^{-3}$ AT$^{-1}$V$^{-2}$) for topological insulator state is one order of magnitude smaller than that ($\sim 1\times 10^{-3}$ AT$^{-1}$V$^{-2}$) for the Weyl semimetal state.}
This again demonstrates the dramatic enhancement of NPHE by Weyl points.

\section{discussion and conclusion}

We have shown that topological band features like Weyl points can help to amplify intrinsic NPHE response. This suggests that topological metals could make suitable platforms for studying intrinsic NPHE. And NPHE measurement may be used as a tool to probe band topology as well as topological phase transitions.

Intrinsic NPHE has a connection to the intrinsic nonlinear Hall effect discussed in magnetic materials with broken inversion symmetry~\cite{gao2014field, wang2021intrinsic, liu2021intrinsic}. A simple picture for intrinsic NPHE is like this: the applied in-plane $B$ field `magnetizes' the electronic band structure and effectively converts a nonmagnetic system into a ferromagnet, then intrinsic NPHE just corresponds to the intrinsic nonlinear Hall effect for this resulting magnetic system. The `magnetizing' effect of $B$ field comes in by coupling to spin and orbital magnetic moments of electrons.
The enhancement by topological band crossing here is
mainly due to the orbital contribution.

It is worth noting that the orbital contribution does not need spin-orbit coupling (SOC), in sharp contrast to the spin contribution, which must require SOC to take action.
This implies that in materials with weak SOC strength, like materials composed of light elements, the response will be dominated by orbital contribution, regardless of band topology. The similar scenario was also reported for linear planar Hall effect~\cite{wang2024orbital}.

This study focuses on intrinsic NPHE. There also exist extrinsic mechanisms for NPHE.
As mentioned, in a simplified picture, NPHE may be related to the $T$-odd nonlinear Hall effect for an effective magnetic system~\cite{gao2014field, wang2021intrinsic, liu2021intrinsic}. The scaling law for $T$-odd nonlinear Hall effect has been proposed in Ref.~\cite{huang2025scaling}, which contains a rich  variety of extrinsic mechanisms. Recently, it was pointed out that for clean systems with long scattering time $\tau$, the leading extrinsic contribution is the so-called Lorentz skew scattering contribution~\cite{xiao2024lorentz}, which originates from the
cooperative action of Lorentz force (from $B$ field) and skew scattering of electrons at disorders.

Our predictions on CuTlSe$_2$ can be directly tested in experiment. To extract nonlinear signal, the standard method is to modulate the driving current with a low frequency (typically less than 100 Hz) and detect the Hall signal at second-harmonic frequency using the lock-in technique. The angular dependence predicted in Eq.~(\ref{angle}) can be probed by etching
CuTlSe$_2$ samples into a disk shape and attaching multiple leads to it to perform measurement at different angles~\cite{kang2019nonlinear, lai2021thirdorder}.
The intrinsic response is independent of scattering. It is usually inferred from a scaling analysis~\cite{huang2025scaling}. This is done by fitting the obtained nonlinear response coefficient $\chi$ by a polynomial function of longitudinal conductivity $\sigma_{xx}\propto\tau$ according to the scaling law. And the intrinsic response is contained in the zeroth order term from such fitting. It should be noted that intrinsic NPHE is a purely Hall response. One needs to remove possible non-Hall (anisotropic resistance) signal in this analysis. This can be done by using the sum-frequency technique, as implemented in a recent experiment~\cite{gao2023quantum}.

In conclusion, we have shown that the intrinsic NPHE can be significantly enhanced by topological nodal features such as Weyl points in a 3D material system. Via a model study, we show that the response exhibits a characteristic resonance lineshape across the Weyl-point energy. This enhancement is mainly via the orbital mechanism, where the in-plane $B$ field couples to the orbital moment of Bloch electrons. This physics is nicely demonstrated in a concrete material example, the chalcopyrite compound CuTlSe$_2$. We show that the different topological states of CuTlSe$_2$ is associated with the degree of tetragonal distortion, and a topological phase transition may be driven by applied strain.
The intrinsic NPHE is found to be much more enhanced in its ideal Weyl semimetal state than topological insulator state.
Our work suggests CuTlSe$_2$ and topological semimetals in general as promising platforms to explore a large intrinsic NPHE, and it also implies  NPHE as a useful tool for probing topological states and topological phase transitions.

\bibliographystyle{apsrev4-2}
\addcontentsline{toc}{section}{\refname}\bibliography{ref}

\end{document}